# Rapid one-step synthesis and compaction of high-performance n-type $Mg_3Sb_2$ thermoelectrics

Jiawei Zhang, Lirong Song, and Bo Brummerstedt Iversen*

**Abstract:** n-type $Mg_3Sb_2$-based compounds are emerging as a promising class of low-cost thermoelectric materials due to their extraordinary performance at low and intermediate temperatures. However, so far high thermoelectric performance has merely been reported in n-type $Mg_3Sb_2$-$Mg_3Bi_2$ alloys with a large amount of Bi. Moreover, current synthesis methods of n-type $Mg_3Sb_2$ bulk thermoelectrics involve multi-step processes that are time- and energy consuming. Here we report a fast and straightforward approach to fabricate n-type $Mg_3Sb_2$ thermoelectrics using spark plasma sintering, which combines the synthesis and compaction in one step. Using this method, we achieve a high thermoelectric figure of merit $zT$ of ~0.4-1.5 at 300-725 K in n-type (Sc, Te)-doped $Mg_3Sb_2$ without alloying with $Mg_3Bi_2$. In comparison with the currently reported synthesis methods, the complexity, process time, and cost of the new method are significantly reduced. This work demonstrates a simple, low-cost route for the potential large-scale production of n-type $Mg_3Sb_2$ thermoelectrics.

As a clean and sustainable solution for the energy crisis, thermoelectric (TE) technology is able to realize the direct interconversion between heat and electricity without moving parts, and it shows great potential in applications such as waste heat harvesting and solid-state refrigeration.[1-5] For the development of this technology, it is essential to have low-cost TE materials that are highly efficient. The performance of TE materials is evaluated by the dimensionless figure of merit $zT = \alpha^2\sigma T/\kappa$, where $\alpha$ is the Seebeck coefficient, $\sigma$ is the electrical conductivity, $T$ is the absolute temperature, and $\kappa$ is the total thermal conductivity.

Recently n-type $Mg_3Sb_2$-based compounds have been intensively studied because of their outstanding TE performance as well as the low-cost and abundantly available constituent elements.[6-18] Exceptional n-type TE properties were first reported in Te-doped $Mg_{3+x}Sb_{1.5}Bi_{0.5}$ with a high $zT$ of ~1.6 at 725 K by Tamaki et al.[10] and Zhang et al.,[9] where the experimental composition and synthesis method reported by Zhang et al. can be traced back to the original work by Pedersen[7] in 2012. Later, several research efforts focused on tuning the carrier scattering mechanism via codoping with the transition metals on the Mg sites and Te on the Sb sites[11,12] or increasing the pressing temperature[9,14,15] in order to improve the carrier mobility at low temperatures. Increasing the pressing temperature to 1123 or 1073 K was found to increase the average grain size, which may reduce grain boundary electrical resistance and thereby improve $zT$ at low temperatures.[19]

Moreover, increasing the Bi content in n-type $Mg_3Sb_2$-$Mg_3Bi_2$ alloys was shown to greatly enhance the room-temperature TE performance,[18,20,21] which can be attributed to the improved weighted mobility due to the reduced conductivity effective mass. Despite significant experimental efforts, superior TE performance with $zT \geq 1.5$ has only been shown in n-type $Mg_3Sb_2$-$Mg_3Bi_2$ alloys. Very few studies have reported on n-type $Mg_3Sb_2$ without alloying with $Mg_3Bi_2$,[8,22-24] and $zT$ is generally below unity. The group-3 elements including La, Y, and Sc have been suggested as effective n-type cation dopants for $Mg_3Sb_2$ in theoretical defect calculations by Gorai et al.,[25] where La and Y have already been confirmed by the experiments.[20,26,27] However, there is no systematic experimental study on Sc doping in $Mg_3Sb_2$. Importantly, so far n-type $Mg_3Sb_2$-based TE materials have been prepared using ball milling,[10-12,14,15,26] arc melting,[9,13,16] or traditional melting[20] followed by spark plasma sintering (SPS), hot pressing, or induction pressing. These methods require multiple steps that are time-, energy-, and cost-intensive, making scalable production challenging.

In this work, we develop a fast direct one-step synthesis and compaction for n-type $Mg_3Sb_2$ TEs using the SPS (see Figure 1a). Compared with the previously reported multi-step synthesis methods, the one-step SPS approach shows a clear advantage that the processing time, energy, and cost are greatly saved. The new approach is demonstrated by fabricating n-type Te-doped, Sc-doped, and (Sc, Te)-codoped $Mg_3Sb_2$ materials without alloying with $Mg_3Bi_2$. n-type Te-doped $Mg_3Sb_2$ prepared by the new method shows comparable or even higher $zT$ values than those reported in Te-doped samples[22-24] prepared by multi-step methods (see Figure 1b). Sc-doped $Mg_3Sb_2$ samples show a maximum $zT$ of ~1.1 at 725 K, superior to Te-doped samples. This indicates the scandium is an effective n-type dopant for $Mg_3Sb_2$, a clear validation for the previous theoretical prediction. An optimal $zT$ as high as ~1.5 at 725 K is obtained in (Sc, Te)-codoped $Mg_3Sb_2$ samples prepared by the one-step SPS (Figure 1b), outperforming any reported n-type $Mg_3Sb_2$ samples[8,22-24] without the $Mg_3Bi_2$ alloying. High $zT$ values of ~0.4-1.5 over a wide range of temperatures from 300 to 725 K in (Sc, Te)-codoped $Mg_3Sb_2$ give rise to a high average $zT$ of ~0.9 (see Figure 1c), approximately two times higher than those of the previously reported Te-doped $Mg_3Sb_2$.[8,22-24]

SPS pressing is one of the commonly used compaction techniques for TE materials. It allows rapid heating and densification of the material within minutes through the joule heat, which is induced by the large pulsed direct current passing through the graphite die/punches as well as the material (see Figure 1a). In this work, the idea is to realize the solid-state reaction and consolidation simultaneously using the SPS. Here $Mg_{3+\delta}Sb_{2-x}Te_x$, $Mg_{3+\delta}Sc_ySb_2$, and $Mg_{3+\delta}Sc_ySb_{2-x}Te_x$ samples with a wide range of compositions were synthesized by thoroughly blending the fine powders of the raw elements followed by the direct SPS pressing at 1073 K (Figure 1a). For clarity, only nine

[*]    Dr. J. Zhang, Dr. L. Song, Prof. Dr. B. B. Iversen
Center for Materials Crystallography
Department of Chemistry and iNANO, Aarhus University
Langelandsgade 140, 8000 Aarhus (Denmark)
E-mail: bo@chem.au.dk

Supporting information for this article is given via a link at the end of the document.

samples are compared in the main text, and all other samples are shown in the Supporting Information.

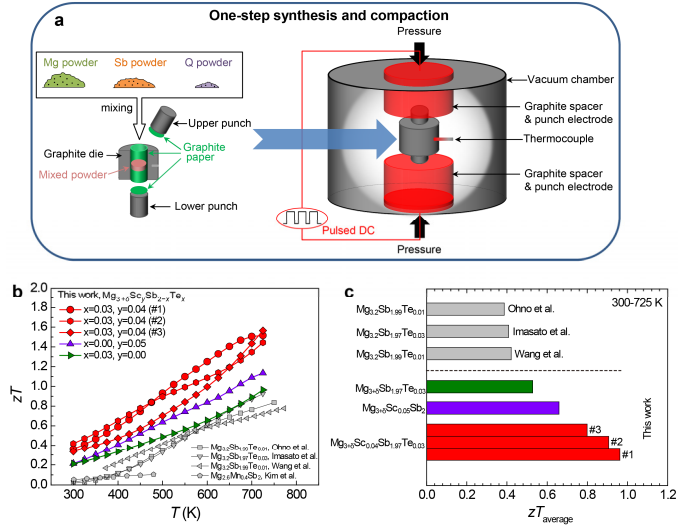

**Figure 1.** a) Schematic diagram of the direct one-step synthesis and compaction of n-type $Mg_3Sb_2$ TE materials using the spark plasma sintering. Q represents the n-type dopant. b,c) Temperature-dependent (b) $zT$ and (c) average $zT$ within the temperature range of 300-725 K for n-type $Mg_{3+\delta}Sc_ySb_{2-x}Te_x$ ($\delta$ = 0.5) prepared by the one-step SPS in comparison with those of the reported n-type $Mg_3Sb_2$ samples[8,22-24] without alloying with $Mg_3Bi_2$.

Figure 2a and Figure S1-S3 in the Supporting Information show powder X-ray diffraction (PXRD) patterns of all SPS-pressed pellets with nominal compositions of $Mg_{3+\delta}Sb_{2-x}Te_x$ ($\delta$ = 0.4-0.6, $x$ = 0.01-0.04), $Mg_{3+\delta}Sc_ySb_2$ ($\delta$ = 0.5, $y$ = 0.02-0.10), and $Mg_{3+\delta}Sc_ySb_{2-x}Te_x$ ($\delta$ = 0.45-0.55, $y$ = 0.03-0.05, $x$ = 0.03). It is well known that the current applied during the SPS pressing might lead to the migration of metal ions and thereby the composition gradient along the pressing direction.[28] Therefore, PXRD patterns of both the top and bottom sides of all pellets were characterized. The results show no clear difference in phase composition between the top and bottom side of each sample. All main peaks in the diffraction patterns can be well indexed to the $\alpha$-$Mg_3Sb_2$ phase with the inverse $\alpha$-$La_2O_3$ structure (space group: $P\bar{3}m1$). The lattice parameters of the top and bottom side of each sample are comparable (Supporting Information, Table S1). Small amounts of MgO were found in all as-pressed pellets, which is likely caused by oxidization during the SPS. Moreover, the Sc-doped samples often show a few very minor unknown impurity peaks between 26 and 33°, which should be related to Sc since Te-doped samples do not show these minor peaks (Supporting Information, Figures S1-S3).

All pellets prepared by the one-step SPS method have relative densities larger than 95% (Supporting Information, Table S2). Figure 2b,c shows scanning electron microscope (SEM) images of the polished surface and fractured cross section of the high-performance $Mg_{3.5}Sc_{0.04}Sb_{1.97}Te_{0.03}$ sample (#1). The result confirms the densified sample with no clear cracks or holes. The actual composition of the $Mg_{3.5}Sc_{0.04}Sb_{1.97}Te_{0.03}$ sample determined by SEM-EDS is close to the nominal value (Supporting Information, Table S3). The lower actual Mg content is typically induced by the evaporation loss of Mg during the SPS due to the high vapor pressure of Mg. The corresponding SEM-EDS mapping images of both the surface and cross section indicate uniform distributions of Mg, Sb, Sc, and Te throughout the sample (see Figure 2d-g and Figure S4 in the Supporting Information).

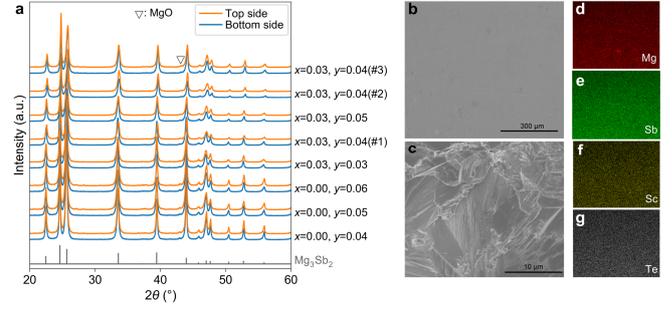

**Figure 2.** a) PXRD patterns of both the top and bottom sides of the pellets with the composition $Mg_{3+\delta}Sc_ySb_{2-x}Te_x$ ($\delta$ = 0.5, $x$ = 0 and 0.03, $y$ = 0.03-0.06). b,c) SEM images of the surface (b) and the fractured cross section (c) of the high-performance sample $Mg_{3.5}Sc_{0.04}Sb_{1.97}Te_{0.03}$(#1). d-g) SEM-EDS elemental mapping images of the surface of the sample $Mg_{3.5}Sc_{0.04}Sb_{1.97}Te_{0.03}$(#1).

The essential feature that contributes to the excellent n-type electrical transport properties in $Mg_3Sb_2$ is the high valley degeneracy $N_v$ = 6 at the conduction band minimum.[9,13] The high valley degeneracy is very beneficial to increasing the density of states effective mass $m_d^*$ and Seebeck coefficient without explicitly decreasing the carrier mobility.[1,2,29-32] As a result, nearly all n-type $Mg_3Sb_2$ samples typically show high room-temperature Seebeck coefficient values larger than ~210 $\mu$V K$^{-1}$ within the carrier density range of ~0.58-1.93 × $10^{19}$ cm$^{-3}$. The Pisarenko plot (Seebeck coefficient *versus* Hall carrier concentration) of n-type $Mg_3Sb_2$ is shown in Figure 3a. Most of the experimental data of n-type (Sc, Te)-doped $Mg_3Sb_2$ samples show reasonably good agreements with the previous theoretical results[9,13] simulated based on the full *ab initio* band structure. Some of the experimental data deviating from theory may be attributed to factors such as the minor impurity phases and doping on the cation sites.[6]

The light conductivity effective mass is another electronic origin for the superior electronic transport in n-type $Mg_3Sb_2$,[9,13] which ensures good carrier mobility. As expected, all n-type (Sc, Te)-doped $Mg_3Sb_2$ samples generally show high room-temperature mobility values ranging from ~51 to ~93 cm$^2$ V$^{-1}$ s$^{-1}$ (see Figure 3b; Supporting Information, Figures S5-S7), which are larger than those of the reported p-type $Mg_3Sb_2$[33] (~16 cm$^2$ V$^{-1}$ s$^{-1}$) and n-type Te-doped $Mg_3Sb_2$[8,23,24] at a comparable carrier concentration. The mobility of all samples generally follows the temperature dependence of $T^{-p}$ (1 $\leq$ p $\leq$ 1.5), indicating a dominant acoustic phonon scattering mechanism. Among all samples, $Mg_{3.5}Sc_{0.04}Sb_{1.97}Te_{0.03}$ (#1) with the best TE performance shows a high mobility of ~76 cm$^2$ V$^{-1}$ s$^{-1}$ at 300 K, which decreases to ~23 cm$^2$ V$^{-1}$ s$^{-1}$ at 725 K. Comparing with many n-type $Mg_3Sb_{1.5}Bi_{0.5}$ and $Mg_3SbBi$ samples,[9,14,15,20] the n-type $Mg_3Sb_2$ samples show slightly lower carrier mobility values but higher Seebeck coefficients at similar carrier concentrations. This can be attributed to the smaller band width and larger

effective mass of the near-edge conduction band in n-type $Mg_3Sb_2$ induced by the wider band gap.[23,34]

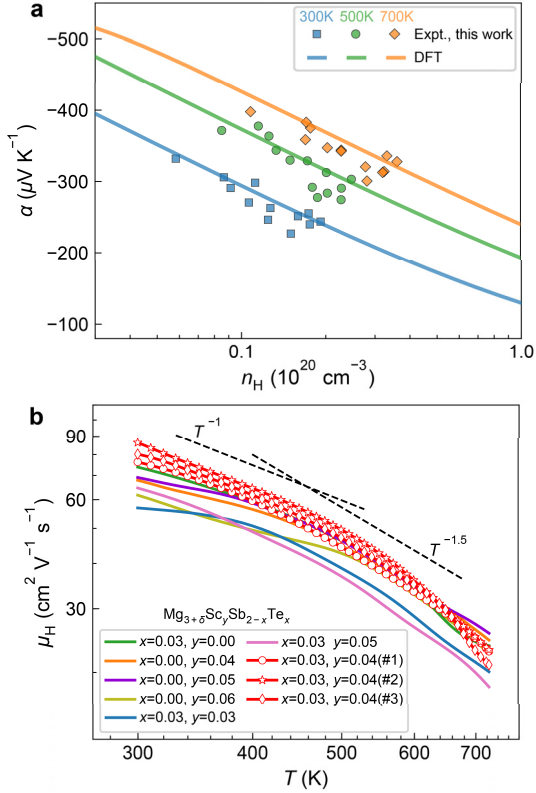

**Figure 3.** a) Seebeck coefficient as a function of Hall carrier concentration for n-type $Mg_{3+\delta}Sc_ySb_{2-x}Te_x$ samples. The solid lines correspond to the theoretical prediction of n-type $Mg_3Sb_2$ based on full band structure with the density functional theory (DFT) calculations taken from Refs. [9,13]. b) Temperature dependence of the Hall mobility of n-type $Mg_{3+\delta}Sc_ySb_{2-x}Te_x$ ($\delta$ = 0.5).

The TE properties of n-type (Sc, Te)-doped $Mg_3Sb_2$ samples prepared by the one-step SPS method are illustrated in Figure 4 and Figures S8-S10 in the Supporting Information. The electrical resistivity and Seebeck coefficient of all Te-doped and (Sc, Te)-co-doped samples show increasing trends with temperature, indicating degenerate semiconductor behaviors. Among the Te-doped samples, $Mg_{3.5}Sb_{1.97}Te_{0.03}$ shows the lowest resistivity of 5.6-9.3 mΩ cm at 300-725 K (Supporting Information, Figure S8a). By introducing a proper amount of Sc, the resistivity can be further reduced to 4.2-7.9 mΩ cm at 300-725 K in $Mg_{3.5}Sc_{0.04}Sb_{1.97}Te_{0.03}$ (Figure 4a). With the combined effect of the low resistivity and moderate Seebeck coefficient, $Mg_{3.5}Sc_{0.04}Sb_{1.97}Te_{0.03}$ shows considerably enhanced power factors ranging from 14.0 to 16.5 $\mu W\,cm^{-1}\,K^{-2}$ within the entire measurement temperature range (Figure 4c), comparable to many reported high-performance n-doped $Mg_{3+\delta}Sb_{1.5}Bi_{0.5}$ samples.[9-12,26]

The temperature-dependent total and lattice thermal conductivity of all n-type samples are displayed in Figure 4d and Figures S8d-S10d in the Supporting Information. The lattice thermal conductivity $\kappa_L$ is obtained by subtracting the electronic thermal conductivity $\kappa_e$ from the total thermal conductivity. The electronic thermal conductivity $\kappa_e$ is evaluated with the Wiedemann-Franz law $\kappa_e = L\sigma T$, where the Lorenz number $L$ is estimated using a single parabolic band model under the acoustic phonon scattering. All n-type $Mg_3Sb_2$ samples generally show reasonably low lattice thermal conductivity values, which, to a large extent, are due to the intrinsic low $\kappa_L$. The intrinsic low $\kappa_L$ in $Mg_3Sb_2$ is mainly related to the large anharmonic scattering rate induced by the soft transverse acoustic phonon modes.[6,35] Comparing with the pure $Mg_3Sb_2$ prepared using a similar synthesis method,[36] the Te doping in $Mg_3Sb_2$ slightly reduces the lattice thermal conductivity. With the proper Sc and Te co-doping, the lattice thermal conductivity at 300 K can be effectively reduced to 1.05 W $m^{-1}$ $K^{-1}$ in $Mg_{3.5}Sc_{0.04}Sb_{1.97}Te_{0.03}$, which decreases to 0.54 W $m^{-1}$ $K^{-1}$ at 725 K (see Figure 4d). The reduced lattice thermal conductivity here is likely due to the enhanced phonon scattering of point defects.

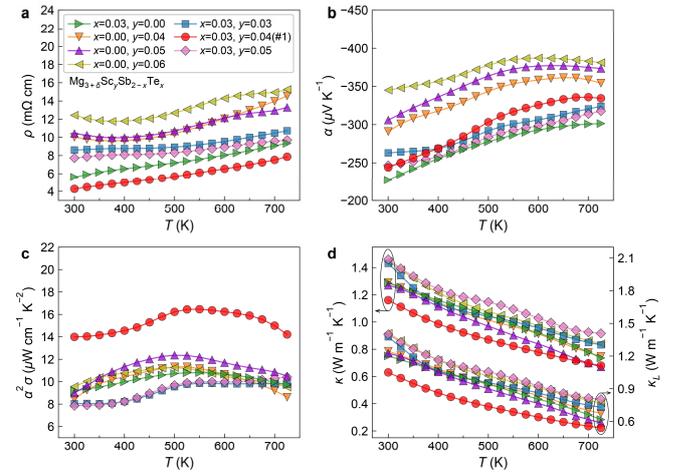

**Figure 4.** Temperature-dependent a) electrical resistivity, b) Seebeck coefficient, c) power factor, and d) total thermal conductivity of n-type $Mg_{3+\delta}Sc_ySb_{2-x}Te_x$ ($\delta$ = 0.5) prepared by the one-step SPS.

The superior power factor in combination with the low thermal conductivity results in excellent overall TE performance in n-type (Sc, Te)-doped $Mg_3Sb_2$. In Te-doped samples, n-type $Mg_{3.5}Sb_{1.97}Te_{0.03}$ shows an optimal zT ranging from 0.21-0.96 at 300-725 K (Figure 1b; Supporting Information, Figure S11a), which is slightly higher than those reported in n-doped $Mg_3Sb_2$ without the $Mg_3Bi_2$ alloying.[8,22-24] Among Sc-doped samples, an optimal zT of 0.21-1.13 at 300-725 K is realized in the n-type $Mg_{3.5}Sc_{0.05}Sb_2$ sample (Figure 1b; Supporting Information, Figure S11b). By co-doping with Sc and Te, zT values can be further enhanced to ~0.4-1.5 at 300-725 K in the n-type $Mg_{3.5}Sc_{0.04}Sb_{1.97}Te_{0.03}$ (Figure 1b; Supporting Information, Figure S11c). Without $Mg_3Bi_2$ alloying, the n-type (Sc, Te)-doped $Mg_3Sb_2$ samples prepared by the one-step SPS method show high zT values comparable to many reported n-doped $Mg_{3+\delta}Sb_{1.5}Bi_{0.5}$ samples.[9-12,15] Figure 5 shows TE properties of three high-performing samples with the same composition $Mg_{3.5}Sc_{0.04}Sb_{1.97}Te_{0.03}$. Considering the measurement uncertainty, the transport properties of the three samples show reasonably good consistency. As a result, high TE performance with zT ≈ 1.5 at 725 K is confirmed in all three n-type

$Mg_{3.5}Sc_{0.04}Sb_{1.97}Te_{0.03}$ samples, indicating good reproducibility. Moreover, the electrical resistivity values are consistent within different thermal cycles (Supporting Information, Figure S12), showing good repeatability.

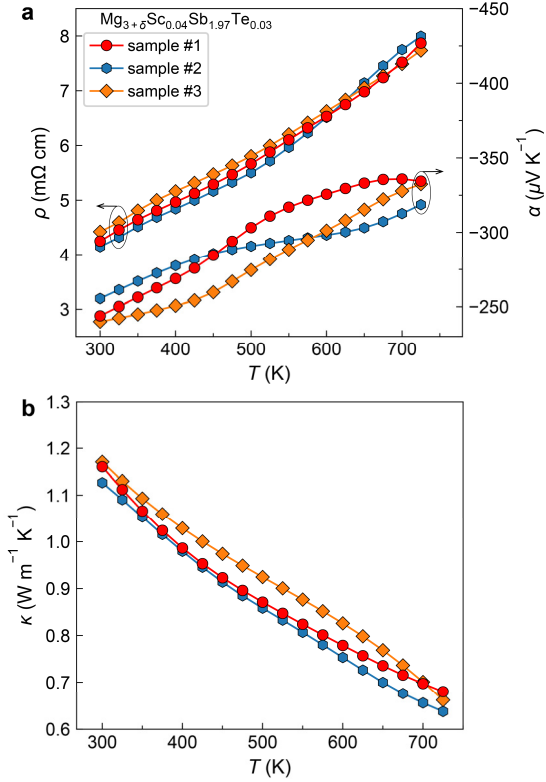

**Figure 5.** Temperature dependence of a) electrical resistivity as well as Seebeck coefficient and b) thermal conductivity of the three high-performing n-type $Mg_{3.5}Sc_{0.04}Sb_{1.97}Te_{0.03}$ samples showing good reproducibility.

In summary, we have demonstrated that a very fast single-step approach can be used to prepare high-performance low-cost n-type $Mg_3Sb_2$ bulk thermoelectrics. This simple approach combines the synthesis and consolidation in one step using spark plasma sintering. We have successfully illustrated this promising method on n-type (Sc, Te)-doped $Mg_3Sb_2$ without alloying with $Mg_3Bi_2$, which shows a very high $zT$ = ~1.5 at 725 K. The TE properties of the bulk samples prepared by the one-step SPS are comparable or even superior to those fabricated by the reported methods. More importantly, the approach is much easier and faster than the reported multistep synthesis methods, and it may be readily extended to the preparation of n-type $Mg_3Sb_2$ with many other n-type dopants. This work provides a new avenue for the rapid, low-cost, and large scale production of n-type $Mg_3Sb_2$ TE materials.

## Experimental Section

*Sample synthesis*: All samples were synthesized and consolidated in one step using the spark plasma sintering (SPS). This method has already been successfully applied in preparing phase-pure TE materials including $Zn_4Sb_3$,[37] ZnSb,[38] p-type Ag-doped $Mg_3Sb_2$,[36] and $Cu_2Se$.[39] Excess Mg is required to compensate the evaporation loss of Mg during the SPS pressing.[40] The amount of excess Mg required to achieve n-type properties might vary with the different sintering atmospheres. High-purity Mg powder (99.8%, -325 mesh, Alfa Aesar), Sb powder (99.5%, -325 mesh, Chempur), Te powder (99.99%, -325 mesh, Alfa Aesar) and Sc powder (99.9%, -100 mesh, Chempur) were weighed according to the nominal compositions of $Mg_{3+\delta}Sb_{2-x}Te_x$ ($\delta$ = 0.4-0.6, $x$ = 0.01-0.04), $Mg_{3+\delta}Sc_ySb_2$ ($\delta$ = 0.5, $y$ = 0.03-0.10), and $Mg_{3+\delta}Sc_ySb_{2-x}Te_x$ ($\delta$ = 0.45-0.55, $y$ = 0.03-0.05, $x$ = 0.03), and then thoroughly mixed in a mixer (SpectroMill, Chemplex Industries, Inc.) for approximately 15 min. The mixed powders were loaded into a half-inch diameter graphite die with graphite paper inserted between the powders and die/punches. The samples were immediately consolidated and sintered in dynamic vacuum under a uniaxial pressure of 60 MPa using an SPS-515S instrument (SPS Syntex Inc., Japan) by heating to 673 K with a 10 min dwell and then ramping to 1073 K for another 4 min. Two $Mg_{3.5}Sc_{0.04}Sb_{1.97}Te_{0.03}$ samples (#2 and #3) were sintered at 1073 K for 10 min. For $Mg_{3+\delta}Sc_ySb_2$ and $Mg_{3+\delta}Sc_ySb_{2-x}Te_x$, powders were weighed, mixed, and loaded into a graphite die inside an argon-filled glove box, whereas these procedures were conducted in air during the synthesis of $Mg_{3+\delta}Sb_{2-x}Te_x$.

*Structure characterization*: Phase purity of the SPS-pressed $Mg_{3+\delta}Sc_ySb_2$ and $Mg_{3+\delta}Sc_ySb_{2-x}Te_x$ pellets was checked by powder X-ray diffraction (PXRD) measured on a Rigaku Smartlab equipped with a Cu K$\alpha_1$ source and the Bragg-Brentano optic. PXRD patterns of the $Mg_{3+\delta}Sb_{2-x}Te_x$ samples were measured using a Rigaku Smartlab equipped with a Co K$\alpha$ source and the parallel beam optic. The lattice parameters of all samples were obtained using the Le Bail method in the JANA2006 program.[41,42] The microstructure and elemental distribution were characterized using a scanning electron microscope (FEI Nova NanoSEM 600) equipped with an energy dispersive spectrometer (EDS).

*Thermoelectric transport property measurements*: The electrical resistivity $\rho$ and Hall coefficient ($R_H$) were measured using the *Van der Pauw* method in dynamic vacuum under a magnetic field up to 1.25 T.[43,44] Hall carrier concentration ($n_H$) and mobility ($\mu_H$) were determined respectively by $1/eR_H$ and $R_H/\rho$, where $e$ is the elementary charge. The pellets coated with a thin layer of BN were annealed by running the Hall measurement for 3 cycles, where the final stabilized cycle is used for discussion in the main text. The Seebeck coefficients of the pellets were then measured with the slope method using chromel-niobium thermocouples in dynamic vacuum on an in-house system,[44] which has a geometry similar to the one reported by Iwanaga et al.[45] The thermal diffusivity ($D$) was measured using the laser flash method (Netzsch, LFA457). Heat capacity ($C_P$) was indirectly estimated using a Pyroceram 9606 standard sample as the reference for the temperature range from 300 to 725 K. The density ($d$) was estimated from the sample mass and volume (Supporting Information, Table S2). Thermal conductivity was then calculated by $\kappa = dDC_P$. In this work, the transport data measured upon cooling were adopted for discussions. The measurement uncertainties of Seebeck coefficient, electrical resistivity, and thermal conductivity are 7%, 5%, and 7%,[46,47] respectively. The combined uncertainty for $zT$ is approximately 20%.

## Acknowledgements

This work was supported by the Danish National Research Foundation (DNRF93). Affiliation with the Center for Integrated Materials Research (iMAT) at Aarhus University is gratefully acknowledged.